\documentclass[twocolumn]{svjour3}

\usepackage{makecell}
\usepackage{graphicx}
\usepackage{amsmath,amssymb,amsfonts}
\usepackage{bm}
\usepackage{color}
\usepackage{tcolorbox}
\usepackage{array}
\usepackage{caption} 
\usepackage{multirow}
\usepackage{adjustbox}
\usepackage[flushleft]{threeparttable}
\usepackage{enumitem}
\usepackage{pbox}
\usepackage{outlines}
\usepackage{caption}
\usepackage{soul}
\usepackage{comment}
\usepackage{setspace}
\usepackage{subfigure}
\usepackage{multirow}

\newcolumntype{M}{>{\centering\arraybackslash}m{0.6cm}}
\newcolumntype{L}[1]{>{\raggedright\let\newline\\\arraybackslash}p{#1}}

\begin{document}
	
	\title{A Preliminary Study on the Potential Usefulness of Open Domain Model for Missing Software Requirements Recommendation}

	\author{Ziyan Zhao,Li Zhang,Xiaoli Lian 
	}
	
	\institute{SKLSDE, Beihang University, Beijing, China, 100191, 
		\email{\{zhaoziyan,lily,lianxiaoli\}@buaa.edu.cn}
	}

\maketitle

\begin{abstract}
Completeness is one of the most important attributes of software requirement specifications. Unfortunately, incompleteness is meanwhile one of the most difficult problems to detect. Some approaches have been proposed to detect missing requirements based on the requirement-oriented domain model. However, this kind of models are lacking for lots of domains. Fortunately, the domain models constructed for different purposes can usually be found online. This raises a question: whether or not these domain models are helpful in finding the missing functional information in requirement specification? To explore this question, we design and conduct a preliminary study by computing the overlapping rate between the entities in domain models and the concepts of natural language software requirements and then digging into four regularities of the occurrence of these entities(concepts) based on two example domains. The usefulness of these regularities, especially the one based on our proposed metric AHME (with $F_2$ gains of 146\% and 223\% on the two domains than without any regularity), has been shown in experiments.
\end{abstract}


\section{Introduction}
\label{sec:intro}

It is extremely common that some basic or even crucial functions are lacking from the software requirement specifications due to the limited domain knowledge of requirements analysts, the short-term plan of companies, and the nature of the change of requirements. However, the functions and capacities that software products can provide constitute their core competitiveness, and the missing of the critical functions seriously decreases their reputation and market share. Therefore, organizations need to make clear their missing functions for sound and practical improvements \cite{zowghi2003interplay}.

Many research focuses on detecting the missing functions based on requirement-oriented domain models \cite{KaiyaQSIC2005,6032492,LionelESEM}. One possible reason that makes the domain model popular in this field is that it is an explicit representation of the salient concepts in an application domain and the relations between these concepts \cite{EvansDomainModel}. And it also depicts the whole picture of the functions in a domain in a compassed and structured way. In addition, Arora et al.\cite{LionelESEM} conducted an empirical study on the use of domain models for the completeness checking of software requirements and showed that the domain model constructed from software requirements is sufficiently sensitive to the information on missing requirements. However, unfortunately, this kind of software requirements-oriented domain model does not exist for many domains. 
 
  Although there are already some studies about the technologies of automatically extracting domain models from software requirements \cite{LionelESEM}, \cite{7582823}, it is not easy to construct the model from scratch due to two problems. Firstly, some tedious manual work with expertise is required for validating and improving the model for practical usage. Second, enough requirements of similar but different systems in the same domain are required for the model construction to guarantee its coverage. Unfortunately, it is not easy to collect enough requirement specifications of multiple competing products, especially for academic researchers, due to the confidentiality and the under-specified requirements in practice.  

Fortunately, some domain models (e.g., ontologies) created for different purposes (i.e., not for requirements validation) from other artifacts than pure software requirements can be found online usually. We call this kind of model as open domain model. We focus on \emph{exploring the feasibility of using an open domain model for finding missing functional information in the software requirement specification in this study}. Another reason behind our selection is that even if the group has the requirements-oriented domain model, it is hard to guarantee its synchronization with requirements because the requirements change constantly. Lots of tedious human work by experts is required to maintain the model. On the contrary,  open domain models are usually updated and actively maintained by many professionals. So, it is possible to select at least one high-quality open domain model for some domains.

Undoubtedly, not all domain models are useful. We definitely need to perform a selection. However, even for the selected domain model, due to the discrepancies between the requirements of one specific system and the open domain model on their perspectives, terminologies usage, and the coverage of functions, there must be an obvious, even huge gap in the coverage of functional information. Intuitively, we are firstly curious about the matching (i.e., overlap) degree between them. In other words, \emph{to what extent that the requirements can be mapped into the open domain model?}

Even if there are certain overlaps, it is easy to expect the gap must be bigger, i.e., much more concepts of domain models that cannot be found in software requirements. Almost all the existing researches on missing function detection propose detecting the concepts that appear in the domain model but not in requirements \cite{LionelESEM,KaiyaQSIC2005,10.1007/978-3-319-77243-1_8}. However, not all missing concepts indicate the missing requirements. For instance, in the Building Automation System, the air-conditioning control system is one necessary piece of equipment that must be installed. However, the heating function, as one child of air-conditioning system in a hierarchical domain model, does not need to be considered in the low latitude area, where the year-round temperature is pretty high. Thus, effectively detecting the appropriate concepts and recommending the valid missing information from the large unmapped scope is essential. Before achieving this ultimate goal, we need to answer one question: \emph{considering the sparse mapping between requirements and domain models, is it possible to find valid clues of missing functional information from the open domain models?}

To be specific, we first build the mapping between the concepts of requirements and domain models based on the synonymous identification and then dig out the occurrence regularities of the mapped concepts from the aspects of requirements and domain models. We consider the distribution characteristics of the requirement entities in the open domain model and requirements as ``regularities''. We concentrate on concept mapping and analysis because it is the general principle of current approaches to detecting missing functional information from requirements relying on any external resources \cite{LionelESEM}. This work is illustrated with two case domains of Unmanned Aerial Vehicle (UAV) and Building Automation System (BAS).

Due to the limited overlap between these two artifacts (i.e., 43.4\% and 20\% for the two cases), the assistance for the missing recommendation must be limited too (i.e., 26\% and 41\% of $F_2$ with the best AHME regularity for the two cases). However, it is well known that there are diverse dependencies between requirements \cite{9218190,pohl1996process,Dahlstedt2005}. Thus, theoretically, the known requirements should be helpful for the unknown requirements recommendation because of their inner semantic associations. 


In order to verify this hypothesis, we complement the open domain model with the already known requirements by aligning them and adding the domain model entity-related requirement concepts as well as the corresponding relationships. Particularly, the recall of recommendations has been greatly increased with almost no loss of precision, although with bigger searching scope and more noise entities, with the gain of 41\% and 72\% for the two cases. 

\textbf{Contribution} The main contribution of this paper includes: 
\begin{itemize}
    \item We propose an approach of automatically building the mapping between NL requirements and the domain model in RDF, including the concepts extraction from NL requirements and the synonyms detection based on external supporting data (i.e., forum). With the case data, we observe that on average, about 31.7\% of requirements can be covered by the open domain model, initially illustrating the usefulness of open domain models on missing functional information recommendation.
    
     \item  We find out four regularities of overlapping entities from both the domain model and requirements perspectives. Experiments show the regularities could effectively reduce the search scope of missing clues in the domain model, making the $F_2$ increase by 13\%-24\% on the two domains.
     
     
    \item We complement the original open domain model with the known requirements through model alignment and relationship deduction, considering the latent semantic associations between requirements. After model complementing, the mapping rates increase to about 73.6\% for the two cases.
    
    \item With model completion, the missing recommendation has been improved further based on our AHME, with the $F_2$ gains of 23\% and 34\%, than on the original domain model with the same regularity and gains of 146\% and 223\% than on the original model without any regularity.

\end{itemize}

The remainder of this paper includes the following sections. Section \ref{sec:relatedWork} introduces the related work on requirements completeness evaluation and missing requirements identification. Section \ref{sec:framework} gives the whole picture of our approach. Sections \ref{subsec: buildMapping}, \ref{subsec: domainModelCompletion} and \ref{subsec: regularityAnalysis} describe the three phases of the detailed procedures, respectively.
Section \ref{sec:usefulness} verifies the effectiveness of our method. Section \ref{sec:discussion} discusses the threats to validity and the limitations of our study. Finally, Section \ref{sec:conclusion} concludes this paper and gives directions to future work.

\section{Related Works}
\label{sec:relatedWork}

Dermeval et al.\cite{dermeval2016applications} conducted a full survey on the applications of domain ontologies in requirements engineering in 2016. They found that although domain ontology has been used in various requirements validation tasks, almost all the researchers built requirements-oriented ontologies, which other researchers seldom used. This is just the primary motivation we explore the open domain model in missing requirements detection.

    Arora et al.\cite{LionelESEM} used three industrial cases to prove that the domain model extracted from requirements is sufficiently sensitive to the missing functions in the requirement set. The mapping relationship between the domain model and requirements needs to be established manually by domain experts in their work. Based on the manual mapping between the concepts in requirements and domain ontology, Kaiya et al.\cite{KaiyaQSIC2005} indicated that the requirements might be incomplete if the concepts of domain ontology cannot be mapped to requirements. 
    
    Kamalrudin et al. \cite{6032492} extracted the critical use case interaction model (the abstract behavior pattern) from requirement texts on the premise of the domain pattern inventory and judged the completeness of requirements by comparing the model with the pattern library. 
    Wei Liu et al.\cite{liu2009service} studied the completeness analysis of requirements for service-oriented computing software. 
    Sergio et al.\cite{espana2009evaluating}, based on the functional model, defined the functional requirements model and the completeness of the requirements model. But the approaches on completeness validation or missing information detection is not involved. On the other hand, Geierhos et al.\cite{geierhos2016complete} are concerned about the existing but incomplete single user requirements, especially the semantically incomplete predicate part. 
    Igor et al.\cite{menzel2010experimental} studied the completeness assessment method of requirements documents (especially the software requirements specifications), focusing on the completeness of information types rather than functional information.

We found that almost all studies are conducted based on the domain ontology constructed from requirements by these researches \cite{dermeval2016applications,farfeleder2011ontology}. Besides, we did not find any research exploring the usage of open domain models to recommend the missing functions in software requirements. 

\section{Overview of This Study}
\label{sec:framework}

This section describes the framework of our study, the research questions, and the cases we selected.

\subsection{The Framework of Our Study}
\label{subsec:framework}

\begin{figure*}[!htbp]
\centering
\includegraphics[width=1\textwidth, clip]{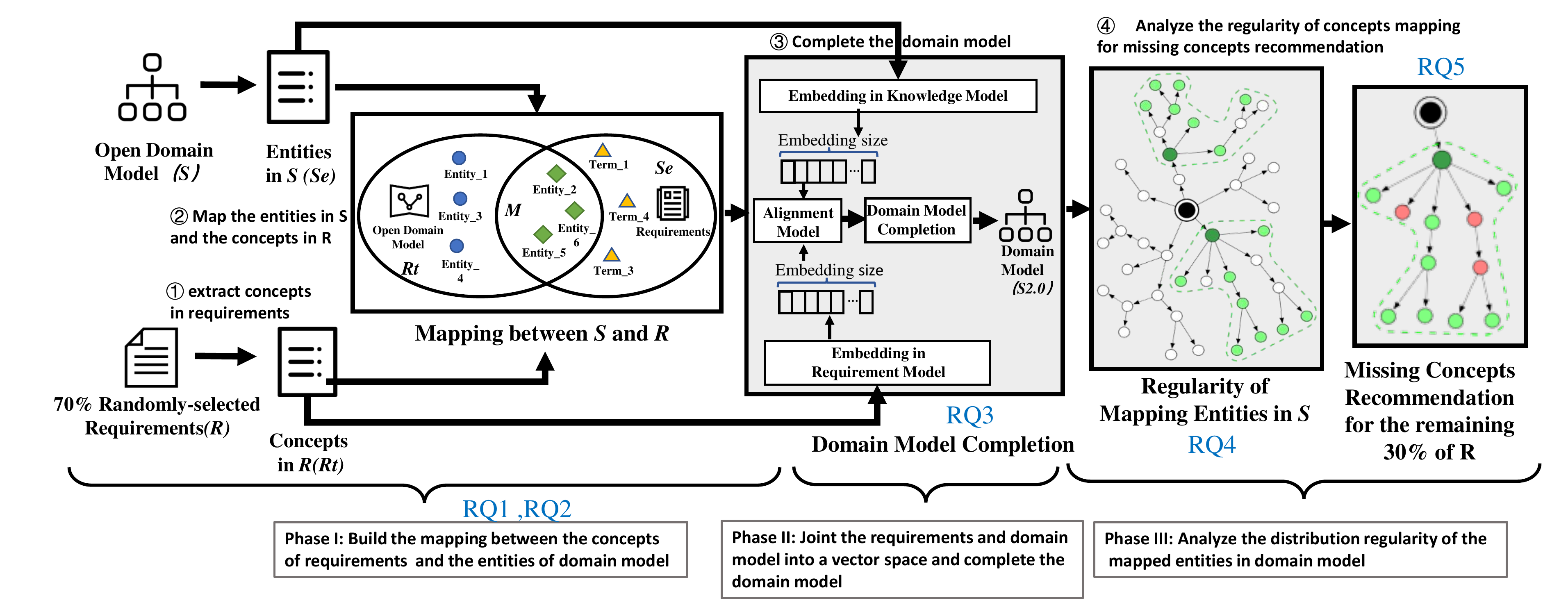} 
\caption{The procedure of our study}\label{fig:framework}
\end{figure*}

Generally, our study includes three phases: mapping construction, domain model completion and regularity analysis. In the first phase, we extract concepts from requirements and construct the mapping between these concepts and the entities in the domain model. In the second phase, we complete the open domain model with the random 70\% requirements through model alignment and relationship deduction. In the third phase,  we analyze the regularity of the occurrence of mapped concepts in both the original and completion domain model for the missing concepts recommendation. The procedure is shown in Fig.~\ref{fig:framework}. 

To be specific, two steps are involved in the first mapping phase. In step one, we randomly select 70$\%$ requirements from the full set (denoted as \emph{R}) and then extract concepts from these selected ones. Then, in step two, we build the mapping relations between these concepts and the entities in the open domain model (denoted as \emph{S}). Note that we only select \emph{Classes entity} from the domain model because we want the conceptual classes that can be mapped with the concepts in requirements.

Then in Phase II, we firstly embed the domain model and requirements with Knowledge Model and Requirement Model. Then align them based on our Alignment Model and complement the domain model with requirements through relationship deduction.

Then in the third phase, we analyze the occurrence of the mapped entities in the domain model from several aspects, including their abstract levels in the model graph, their distributions, and other characteristics. These regularities are expected to narrow down the scope of missed concepts recommendation in requirements because not all of the missing concepts are valuable. 

Finally, based on our regularity, we give our initial ideas about the missing information recommendations and use the 30\% requirements to verify our method. Due to the stochastic elements during the 7:3 partition, we perform the whole process (i.e., Phase I-III) 30 times and calculate the average values on all measurements, including the mapping rate and all recommendation metrics. 

\subsection{Research Questions}
\label{subsec:RQs}

\textbf{RQ1: What is the overlap ratio between the open domain model and software requirements?} To study the degree of association between requirement specifications and the open domain model, we propose using a mapping rate to indicate whether there is a relationship between them and the degree of association. We think it is an essential and vital sign to help to judge whether the domain model is useful for the requirement improvements. 

\textbf{RQ2: To what extent the mapping relationship between open domain model and software requirements can be built automatically?} It is tedious to build the mapping relation manually. Therefore, we propose a method of automatically building mapping relation and comparing it with manual construction.

 \textbf{RQ3: To what extent that the 70\% requirements can complement the  domain model to increase the gap between open domain model and the requirements of one specific system?} Considering the limited overlap between the open domain model and software requirements, we would like to complete the domain model with the known requirements and explore the extent of the completion for the overlap increase.
 
\textbf{RQ4: Are there any distribution regularities of the overlapped entities which are potentially useful for the missing requirements identification?}  We primarily focus on three aspects: 1) the entity types in the domain model (i.e., \emph{Classes}, \emph{Properties} or \emph{Name Individuals}); 2) the position of the entity concept (i.e., root, intermediate or leaf node); and 3) the local clusters of conceptual entities in the domain model. We attempt to detect the focus of requirements by analyzing its covered scope in the domain model, expecting to narrow down the scope of missing function recommendations.

\textbf{RQ5: To what extent can the regularities be used as hints about missing requirements recommendation?} We would like to evaluate whether the regularity can be used for missing requirements recommendation. In other words, to what extent that the missing requirements conform to the regularities?

\textbf{RQ6: To what extent can the missing requirements recommendation be improved through the completion domain model?} We would like to explore the effectiveness of the domain model completion on the missing requirements recommendation with our proposed regularities.


\subsection{Introduction of the Case Domains}
\label{subsec:cases}

Note that the quality of open domain models must be varied significantly and we can't just use a random one. Actually, we make an easy filtering. We need requirement sets and the corresponding domain models, which almost depict the same perspective of the domain with requirements (e.g., functions, concepts), although its scope can be much more significant. We obtain all of the objects of our study via public search engines (mainly Google).

It is much easier to collect software requirements than a domain model online. Thus, we prepare a set of software requirements first and then search for the available domain models one by one. There are a few excellent open repositories of software requirements, such as the creation from the group of Center of Excellence for Software \& Systems Traceability (COEST)\footnote{http://coest.org/datasets} and the RE resources collected by the requirements engineering group at the University of Technology, Sydney (RE@UTS)\footnote{http://research.it.uts.edu.au/re/cgi-bin/resources\_srs.cgi}. We download and scan each requirement document in these two repositories to learn the requirements' perspectives and obtain the basic domain search query. Then we attempt to find appropriate domain models via Google by the search queries like \emph{``domain model'' OR ontology OR ``knowledge graph''} AND basic domain query (such as ``drone''). In this paper, we refer to \emph{Ontology}, \emph{Domain Model}, and \emph{Knowledge Graph} as Domain Model. For the convenience of processing, we use Resource Description Framework (RDF)\cite{w3.org/RDF/}, a data model commonly used in the knowledge graph, to represent the domain model \cite{ji2020survey}. Finally, we find 12 models which can be matched with the existing software requirements.

We select two domains for this study based on three considerations. 1) We should be familiar with the domains to evaluate the concepts mapping and the later missing functions recommendation. 2) The domain model should be related to software system development and should be close to the software requirements as much as possible. Moreover, 3) more complete domain models would be selected if there are multiple versions of the same model.  

The two case studies for this current study are Unmanned Aerial Vehicle (UAV, noted as Case A) and Building Automation System (BAS, noted as Case B). 

Case A: The open domain model\footnote{http://www.dronetology.net/} of UAV includes 400 entities and 146 \emph{Classes} entities. The requirements are from the University of Notre Dame\footnote{https://dronology.info/}\cite{cleland2018dronology} include 99 functional requirements. 

Case B: The open domain model of BAS\footnote{https://gitlab.fi.muni.cz/ xkucer16/semanticBMS} includes 484 entities, all of which are \emph{Classes} entities. Moreover, the requirements are from the Standard Building Automation Service (BAS) Specification(2015) \cite{BASReq} consisted of 456 functional requirements. 

We randomly select 70\% (i.e., 70 requirements in Case A and 320 requirements in Case B) for the mapping construction and the regularity analysis. And the remaining 30\% (i.e., 29 requirements in Case A, 136 requirements in Case B) are used in the initial usefulness evaluation of the regularity for missing functional recommendation.

During the process of data collection, we build an initial repository of 117 open models. We public these models as well as the 12 suites of domain model and software requirement descriptions via Github
 \footnote{https://github.com/ZacharyZhao55/Open-Domain-Model}.

\section{Phase I: Mapping Construction}
\label{subsec: buildMapping}

According to Fig.~\ref{fig:framework}, during phase I, we mainly perform two tasks, including salient concepts extraction from NL requirements and domain models, and the mapping construction between domain and requirements concepts.

\subsection{Extracting Concepts and Entities}
\label{subsubsec:TermsExtraction}

Open domain models usually include entities of different categories. In OWL ontology language rules, common categories include ``\emph{Classes,}'' ``\emph{Object Properties,}'' ``\emph{Named Individuals,}'' and so on. To build a mapping of conceptual entities between the domain model and requirements, as shown in the Fig.~\ref{fig:framework}, we extract all of the entity concepts in the ``\emph{Classes}'' of the domain model and represent them with ClassesTerms ($\mathit{CT}$). Besides, we need to extract the concepts from the NL requirements. We use the terms extraction method for this task because of the similar semantic between ``term'' and ``concept'', according to their definitions \cite{sager1980english}, \cite{feng1997introduction}, \cite{Terminologywork2000}. We denote the concepts extracted from requirements as RequirementTerms ($\mathit{RT}$).

It is easy to filter and obtain the entity concept from the domain model by retrieving its topology. We are more focused on the concepts extraction from requirements.

Briefly, there are three commonly used terms extraction methods from the text: rule-based, statistics-based, and the combination of these two kinds of approaches. The rule-based approaches usually explicitly build rules based on linguistic knowledge such as Part-of-Speech (POS) and lexical patterns. Due to the relatively full analysis of the similar domain corpus, these approaches tend to present significantly better extraction accuracy. However, their generality is poor, and most of them are tightly coupled with domains, corpus, and languages \cite{yuan2015survey}. Generally, the statistics-based approach explores the distribution statistical properties of terms in the corpus with statistical theory guidance, such as frequency \cite{vivaldi2007evaluation,park2002automatic}, TF-IDF\cite{bolshakova2013topic}, Domain Relevance and Domain Consensus\cite{velardi2001identification}, Mutual Information\cite{daille1994study}, and Log-likehood\cite{gelbukh2010automatic,cohen1995highlights}. These methods have strong universality and are not limited to specific domains or corpus. However, the reliability of the analysis largely depends on the quality of the corpus. In this work, we select one hybrid approach because it has the advantage of high accuracy and domain independence meanwhile, the advantages of the above two kinds of approaches \cite{yuan2015survey}. To be specific, we use the C-Value method \cite{frantzi2000automatic}, a widespread and effective way for term extraction, which elegantly combines grammatical rules and statistical information. 

Briefly, C-Value selects the Multi-Word Terms (\emph{MWT}) from the corpus through two steps. Firstly, it obtains all candidate terms based on a series of linguistic filters for the nested noun selection, such as ${{Noun}^+{Noun}}$, as well as a stop-word filter. Then it assigns a termhood measure to all candidate strings by considering their total frequency of occurrence, their frequency as part of other longer candidate terms, the number of the longer candidate terms, and their length (i.e., words). The final terms whose termhood is above a predefined threshold would be selected as the concepts of requirements.

 As shown in Table.~\ref{tab:reqTerms}, we extracted concepts from 70$\%$ of the requirements of the two cases. For the 70 requirements of UAV, 77 concepts are extracted automatically, and 53 ones are remained after manual checking. For the 320 requirements of BAS, 167 concepts are automatically extracted, and 145 ones are kept after manual checking. Most of the filtered terms are incomplete concepts with only part of semantic.

\begin{table*}[!htbp]
\centering
\caption{Number of concepts extracted from NL requirements}
\label{tab:reqTerms}
\begin{tabular}{c|c|c|c}
\hline
\multicolumn{1}{c|}{Case} & \begin{tabular}[c]{@{}c@{}} \#Concepts of auto-extraction\end{tabular} & \begin{tabular}[c]{@{}c@{}}\#Concepts after manual checking\end{tabular} & \multicolumn{1}{c}{Examples}                                                                                                              \\ \hline
UAV                        & 77                                                               & 53                                                                & \begin{tabular}[c]{@{}l@{}}UI Middleware, Object\\ Avoidance System, Flight\\ Pattern, Takeoff Command\end{tabular}                       \\ \hline
BAS                        & 167                                                              & 145                                                               & \begin{tabular}[c]{@{}l@{}}Air Conditioning, Building\\ Controller, Advanced \\ Application Controllers,\\ Building Control Unit\end{tabular} \\ \hline
\end{tabular}
\end{table*}

\subsection{Mapping Requirement Concepts with Domain Entities}
\label{subsubsec:Mapping Phase}

After the concepts and entities are extracted from the requirements and domain model, we need to build the mapping between them. 

Two steps are involved. Firstly, we compare the names of entities and terms and build a direct mapping between them. However, synonyms can be commonly found due to the diverse scope and background of the open domain model and software requirements. Thus, we also design to detect the synonymous relationship as a supplement to step one. Due to the limited space, we put more words on the synonymous detection-based mapping construction here.

Considering the sparse semantic information in both software requirements and the domain model, we need to look for lots of external domain data to bridge the synonymous relationship. So, we crawl domain documents extensively online from Google by combining the domain name and the high-frequency terms and entities of requirements and domain model. Finally, ten documents are sorted out from the retrieved results for each of the cases, with the average size of each corpus being 1000 sentences and 25,000 words.

When obtaining the domain corpus, we first extract Multi-Word Terms from them using the above approach of section \ref{subsubsec:TermsExtraction}. Then, the terms from the domain corpus, the requirements, and domain models are represented with embedding vectors. Finally, we calculate the similarity between them \cite{glavavs2016unsupervised}.

Most domain concepts are composed of multiple terms. However, most of the research on term similarity focus on individual words \cite{hazem2018word}, and very little work can be used for synonymous Multi-Word Terms detection. Fortunately, we find one good approach of Hazem \cite{hazem2014semi} which detects the synonymous relation between \emph{MWT} with the same head or tail parts based on the similarity calculation on word embeddings. However, Hazem's \emph{MWT} model of two parts of \emph{head (E)} and \emph{tail (T)} doesn't work for longer terms. Thus, we make a simple extension by adding \emph{middle (M) }element to their definition. 

\par
\begin{align}
MWT=(E;M;T)
\end{align}

In line with our \emph{MWT} definition, we define four basic inference rules for synonymous extraction. Let two terms be $MWT_{1} = (E_{1}; M_{1}; T_{1})$ and $MWT_{2} = (E_{2}; M_{2}; T_{2})$, and $syn(MWT_{1}; MWT_{2} )$ be a synonym relation between them, the inferences rules can be formulated as:

\begin{center}
\fontsize{8pt}{2.5pt}
\begin{tcolorbox}[colback=white,
                  colframe=black,
                  width=8cm,
                  arc=1mm, auto outer arc,
                  boxrule=0.5pt,
                 ]
                $ \bm{R_{1}:}T_{1}=T_{2}\neq\varnothing \wedge M_{1}=M_{2} \wedge syn(E_{1},E_{2} ) \wedge E_{1} \neq\varnothing  \wedge  E_{2} \neq\varnothing \supset syn(MWT_{1},MWT_{2}) \nonumber$
                        	
                $ \bm{R_{2}:}E_{1}=E_{2}\neq\varnothing \wedge M_{1}=M_{2} \wedge syn(T_{1},T_{2} ) \wedge T_{1} \neq\varnothing  \wedge  T_{2} \neq\varnothing \supset syn(MWT_{1},MWT_{2}) \nonumber$    
                        	
                $ \bm{R_{3}:}E_{1}=T_{2}\neq\varnothing \wedge M_{1}=M_{2} \wedge syn(T_{1},E_{2} ) \wedge T_{1} \neq\varnothing  \wedge  E_{2} \neq\varnothing\supset syn(MWT_{1},MWT_{2}) \nonumber$
                        	
                $ \bm{R_{4}:}E_{2}=T_{1}\neq\varnothing \wedge M_{1}=M_{2} \wedge syn(T_{2},E_{1} ) \wedge T_{2} \neq\varnothing  \wedge  E_{1} \neq\varnothing\supset syn(MWT_{1},MWT_{2}) \nonumber$
\end{tcolorbox}
\end{center}

The above four rules follow a straightforward principle. Once two parts of the two Multi-Word Terms are equal (one of them can be empty), and the remaining ones are synonymous, the two terms are synonyms. The synonymous relation between a single word in the three parts is determined based on the general dictionary, like WordNet. Taking the first $R_1$ as an example, given both the heads and tails of two \emph{MWTs}, if they are not empty, and their heads and middle parts are equal respectively, the heads are synonymous in a general dictionary. We consider these two \emph{MWTs} are synonyms. The rules can be extended. For instance, the three parts can be represented as three finer elements for very complicated terms. 

Like most word embedding-based similarity calculation approaches, the terms that are not synonymous but occurred with high frequency are inevitably selected. Therefore, manual screening is required.

Our synonyms detection results on the terms of 70\% requirements and entities of the domain model are shown in Table.~\ref{tab:synonyms}. For the UAV case, there are 14 pairs of synonyms automatically extracted, and ten pairs were kept after manual checking. And for the other case, 11 pairs were automatically extracted, and eight ones were kept.

\begin{table*}[!htbp]
\centering
\caption{Number of Synonyms Extracted in Our Two Cases}
\label{tab:synonyms}
\begin{tabular}{l|c|c|l}
\hline
\multicolumn{1}{c|}{Case} & \begin{tabular}[c]{@{}c@{}}\#Synonym Pairs of\\ Auto-extraction\end{tabular} & \begin{tabular}[c]{@{}c@{}}\#Synonym Pairs After\\ Manual Checking \end{tabular} & \multicolumn{1}{c}{Examples}                                                                                                                       \\ \hline
UAV                        & 14                                                                                 & 10                                                                         & \begin{tabular}[c]{@{}l@{}}\{takeoff altitude, home altitude\}\\  \{flight pattern, flight phase\}...\end{tabular} \\ \hline
BAS                        & 11                                                                                 & 8                                                                          & \begin{tabular}[c]{@{}l@{}}\{alarm panel, alarm console\}\\ \{pressure gauge, pressure rating\},...\end{tabular}                                       \\ \hline
\end{tabular}
\end{table*}

Based on the results of term extraction and synonyms detection, we can build the mapping between the terms of 70\% requirements and the entities of the domain model and calculate the mapping rate. The results are shown in Table.~\ref{tab:mappingRate}. 

\textbf{Addressing RQ1:} For the UAV case, 53 terms were extracted from 70 requirements, of which 23 mapped entities in the domain model, 43.4\% of the mapping rate. For the BAS case, 145 terms were extracted from 320 requirements, and 29 ones can be mapped with the entities of the domain model, with a mapping rate of 20\%. The mapping rate of BAS is lower than that of UAV. The main reason is that the scope of BAS is more significant than that of UAV. We found that the abstract nodes in the UAV domain model accounted for 5.4\% of the Classes, while the abstract nodes in BAS accounted for 8.68\%. We believe that the number of abstract nodes is proportional to the scope of the domain model. Since the functional scope of requirements usually is small, it makes sense that the number of mapped entities of BAS is less than that of UAV.

\begin{table*}[!htbp]
\centering
\caption{Mapping Results of the 70\% Requirements with Open Domain Model}
\label{tab:mappingRate}
\begin{tabular}{c|c|c|c}
\hline
Case & \begin{tabular}[c]{@{}c@{}}Terms in 70\% REQs\end{tabular} & \begin{tabular}[c]{@{}c@{}}Mapped Entities\end{tabular} & Mapping Rate \\ \hline
UAV  & 53                                                               & 23                                                                       & 43.4\%       \\ \hline
BAS  & 145                                                              & 29                                                                       & 20\%         \\ \hline
\end{tabular}
\end{table*}

To address RQ2, we compare a set of automated approaches on the mapping construction, shown in Table.\ref{tab:MappingRateInDifferentApp}. To evaluate the performance of our approach (i.e., C-Value+synonym detection), we select the naive Nonus extraction (including Nouns(NNs) and Noun Phrases(NPs)) and mapping, term extraction-based mapping (pure C-Value) as the baselines. The simple Nouns extraction is selected because it is commonly used in concepts extraction \cite{harmain2003cm}. And we compare the results of pure C-Value to prove the necessity of using entity concepts and synonym extraction methods. The results of these three automated approaches are compared with those of manual constructions.

\begin{table*}[!htbp]
\centering
\caption{Mapping Rate of Different Approaches}
\label{tab:MappingRateInDifferentApp}
\begin{tabular}{c|c|c|c|c}
\hline
Case  & Methods & \begin{tabular}[c]{@{}c@{}}\#Entities in REQ\end{tabular} & \begin{tabular}[c]{@{}c@{}}\#Mapped Entities\end{tabular} & \begin{tabular}[c]{@{}c@{}}Mapping Rate\end{tabular} \\ \hline
\multirow{4}{*}{UAV}    & NNs,NPs   & 135    & 15    & 11.11\%   \\ \cline{2-5} 
 & C-Value     & 53   & 13   & 24.53\%  \\ \cline{2-5} 
 & \begin{tabular}[c]{@{}c@{}}\textbf{C-Value+ Synonym}\end{tabular} & \textbf{53}   & \textbf{23}   & \textbf{43.40\% }  \\ \cline{2-5} 
 & Manual   & 56  & 27  & 48.21\%   \\ \hline
\multicolumn{1}{c|}{\multirow{4}{*}{BAS}} & NNs,NPs   & 452    & 12    & 2.65\%     \\ \cline{2-5} 
\multicolumn{1}{c|}{}   & C-Value & 145    & 21   & 14.48\%   \\ \cline{2-5} 
\multicolumn{1}{c|}{}                     & \begin{tabular}[c]{@{}c@{}}\textbf{C-Value+ synonym}\end{tabular} & \textbf{145 }  & \textbf{29} & \textbf{20\% }  \\ \cline{2-5} 
\multicolumn{1}{c|}{}                     & Manual                                                     & 150                                                                      & 38                                                                      & 25.33\%                                                \\ \hline
\end{tabular}
\end{table*}

\textbf{Addressing RQ2:} The column of \emph{Mapping rate} in Table.\ref{tab:MappingRateInDifferentApp} shows that our approach of the combination of C-Value-based direct mapping and synonymous-based mapping yields much better results than all of the baselines (highlighted with bold font) for both two domains. By comparing the results of \emph{C-Value}-based mapping and the combination with synonymous-based mapping, we find the mapping rates increase by 18.87\% and 5.52\% respectively for the domains of UAV and BAS, indicating that our synonymous extraction-based entities mapping is a solid supplement to the term-based indirect mapping.
Meanwhile, we can see that the approach of NNs and NPs extraction can obtain the most significant number of concepts. However, its mapping rate is noticeably lower than the other approaches because it generated lots of short simple nouns and missed some multi-word terms, which are very likely the concepts.

\section{Phase II: Domain Model Completion}
\label{subsec: domainModelCompletion}

According to Fig.~\ref{fig:framework}, in Phase II, we jointly embed the domain model and requirements into the same vector space and then complete the domain model by merging parts of the entities in requirements and constructing the relationships between them and those already in the domain model.

\subsection{Two Assumptions about the Domain Model and Requirements}

With regards to the coverage of requirements, there may be two conditions about the open domain model.

\textbf{Condition I}: \textbf{The entities are complete and some relationships between entities may be missing.} To be specific, all the concepts in requirement can be found in the domain model, either the exact same expressions or synonyms identified in phase I. But there may be extra relationships between some entities in requirements, but not in the domain model.


\textbf{Condition II}: \textbf{Some entities may be missing.} To be specific, some concepts in requirements are lacking in the domain model. In this case, the corresponding relationships between these concepts must be missing too.

Requirements can be used to complete the domain model for the above two conditions.

\subsection{Joint Embedding}
\label{subsec:embedding}

We embed the open domain model and requirements into the same continuous vector space based on the approach of Huaping at el.\cite{zhong2015aligning,wang2014knowledge}. We select this approach because it does not rely on any external data resource and has good performance on the data sets of different sizes \cite{zhong2015aligning,wang2014knowledge}.

Generally, both the domain model and requirements are modeled firstly as fact triples (i.e., \emph{knowledge model} and \emph{Requirement model}) respectively. Then we adjust their vectors to align them into the same vector through a \emph{alignment model}.

\textbf{Knowledge Model} The domain model is modeled as a set of fact triplets \emph{(h, r, t)}, where $h,t \in \mathcal{E} $(the set of entities) and $r \in \mathcal{R}$(the set of relations), depicting the relation \emph{r} between \emph{h} and \emph{t}. The whole set of the fact triplets in a domain model is annotated as $\bigtriangleup$. The conditional probability of a fact $(h, r, t)$ is defined as:


\par
\begin{small}
\begin{align}
Pr(h|r,t)=\frac{exp\{z(h,r,t)\}}{\sum_{\tilde{h}\in I}^{}exp\{z(\tilde{h},r,t)\}}
\end{align}
\end{small}

in \cite{zhong2015aligning,wang2014knowledge}, where $z(h,r,t) = b-\frac{1}{2}\left\|h+r-t \right\|^{2}$, and $b$ is a constant for bias designated for adjusting the scale for better numerical stability. The model both defines $Pr(r|h,t)$ and $Pr(t|h,r)$ in the same way with normalization respectively. The likelihood of observing a fact triplet is defined as:

\par
\begin{small}
\begin{align}
\mathcal{L}_{f}(h,r,t)=log Pr(h|r,t)+log Pr(r|h,t) + log Pr(t|h,r)
\end{align}
\end{small}

The goal of the knowledge model is to maximize the conditional likelihood functions of the existing triples $(h, r, t)$ in the domain model:

\par
\begin{small}
\begin{align}
\mathcal{L}_{K}(h,r,t)=-\sum_{(h,r,t)\in\bigtriangleup}^{} \mathcal{L}_{f}(h,r,t)
\end{align}
\end{small}

\textbf{Requirement Model} If two terms \emph{w} and \emph{v} co-occur in a requirements, we assume that there is a relationship $r_{wv}$ between \emph{w} and \emph{v}, and we use $(w,r_{wv},v)$ to represent the relation between \emph{w} and \emph{v}. The set of terms in requirements includes the MWT extracted in Section \ref{subsubsec:TermsExtraction} and also the remaining nouns which appear as unique semantic units in any requirement or window. We annotate the term set as $\mathcal{V}$.

The probability of term pair \emph{w} and \emph{v} co-occurring in a requirement can be defined as:

\par
\begin{center}
\begin{small}
\begin{align}
Pr(w|v)=\frac{exp\{z(w,v)\}}{\sum_{\tilde{w}\in \mathcal{V}}^{}exp\{z(\tilde{w},v)\}}
\end{align}
\end{small}
\end{center}

Then the loss function of requirement model is 

\par
\begin{small}
\begin{align}
\mathcal{L}_{R}(w,v)=-\sum_{(w,v)\in\bigtriangleup}^{} logPr(w|v)
\end{align}
\end{small}

\textbf{Alignment Model} The alignment is performed based on the same expression or synonymous relationship between the entities in domain model and the terms in requirements. To be specific, for a fact triplet $(h,r,t) \in \bigtriangleup$, if the name of entity \emph{h} equals with or is synonymous with $w_h \in \mathcal{V}$ according to the results in Section \ref{subsubsec:TermsExtraction}, then generate a new triplet $(w_h,r,t)$. Similarly, if the names of entity \emph{t} and $w_t$ are the same or synonymous and $w_t \in \mathcal{V}$, generate $(h,r,w_t)$ and $(w_h,r,w_t)$. The loss function of alignment model is

\par
\begin{small}
\begin{align}
\mathcal{L}_{A}=&-\sum_{(h,r,t)\in\bigtriangleup}^{} I_{[w_h \in \mathcal{V}]\wedge w_t \in \mathcal{V}} \cdot \mathcal{L}_f(w_h,r,w_t) \nonumber\\&+I_{[w_h \in \mathcal{V}]} \cdot \mathcal{L}_f(w_h,r,t) + I_{[w_t \in \mathcal{V}]} \cdot \mathcal{L}_f(w,r,w_t) 
\end{align}
\end{small}

Considering the above three parts, the likelihood function of the joint learning model can be defined as:

\par
\begin{small}
\begin{align}
\mathcal{L}= \mathcal{L}_{K} +\mathcal{L}_{R} +\mathcal{L}_{A}
\end{align}
\end{small}

Before the alignment, we initialize the vectors of fact triplets in the knowledge model and requirement model with random integers generated by a uniform distribution. Overall, there are 1567 triplets in the UAV domain model, consisting of 400 entities and 14 relationship types. And for the BAS domain model, there are 493 triplets with 484 entities and only one \emph{subClassOf} relationship. And there are 137 and 958 terms in 70\% of UAV and BAS requirements.

\subsection{Domain Model Completion}
\label{subsec:modelCompletion}

The target of domain model completion is to add the concepts in requirements satisfying the following two conditions to the domain model meanwhile: 1) the concepts appear in requirements but not in the domain model, and 2) each of them has parent-child relationship (usually \emph{hasSubClasses} or \emph{subClassOf})  with at least one  entity in domain model. To achieve this goal, we need to explore the relationship types between these requirement concepts and the domain entities, and also between the newly added requirement concept and the existing added ones.

According to Section \ref{subsec:embedding}, we can infer that in a fact triplet $(h,r,t)$, the relation vector $r$ can be represented as $r=(t-h)$. Considering that after alignment, each entity of requirements and domain models has been assigned a vector, we can infer the relationships to be added based on this triangular law.

However, through the vector subtraction, we can only infer that there may be certain relationship between two entities. We can't know the relationship types and they are critical for the missing requirement recommendation.

We assume that the angles between the pairs of two entities with the same relationship should be the same. Let the same relationship exist between entities \emph{a} and \emph{b}, \emph{c} and \emph{d}. The cosine value of the angles between $\Vec{a}$ and $\Vec{b}$, between $\Vec{c}$ and $\Vec{d}$ should be equivalent. With this assumption, we identify the requirement entities with required parent-child relationships with the domain model entities or already added requirement entities.

Obviously, the added entities and relationships will change the structure of the original domain model. Especially, lots of ``root'' nodes will be identified, violating the original ``one-root'' top-to-bottom design of the domain model. Therefore, we make a compromise by regarding all added entities as the children of some related domain model entities. To be specific, for a to-be-added requirement entity \emph{Re}, which is the parent of a domain model entity \emph{De} according to our computing,
\begin{itemize}
    \item if it doesn't have any parent-child relationship with other requirement entities, we add it as one child of \emph{De}.
    \item if it is the parent of an requirement entity \emph{Re'} which is also the child of \emph{De}, we adjust the original structure of $Re \to De \to Re'$ to $De \to Re \to Re'$.
\end{itemize}

In the original UAV domain model, the relationship number is 1567, including 123 \emph{hasSubClasses} relationships. After completion, 311 \emph{hasSubClasses} relationships are supplemented, making the total of 434.
For BAS, there are 493 \emph{subClassOf} relationships in the original domain model, and 641 \emph{subClassOf} relationships are supplemented after completion, making the total relationships of 1135.

While obtaining the complemented domain models, we recalculate the number of mapped entities between the requirements and domain model, as shown in the Table.~ \ref{tab:MappingRateInCompletionDomainModel}. The mapping is also performed by the C-Value+Synonym method in Section \ref{subsec: buildMapping}. Considering the stochastic performance of the alignment model, we run the model alignment + completion 30 times and obtain the average values as the mapping results.

We can observe that the rate of the mapped entities in all entities is about 75.47\% and 71.72\%, increasing by 30\%-50\% compared with the results in Table.~\ref{tab:mappingRate}. Meanwhile, we can see that neither of them reaches 100\%, which means that some entities in requirements cannot be mapped to the domain model due to the missing relationships between them and the domain model entities. Besides, we are clear that a higher mapping rate does not mean better recommendations because of the unclear associations between the entities in the 70\% requirements and the remaining 30\% ones. So the further experiment is critical.

\begin{table*}[!htbp]
\centering
\caption{Mapping Results of the 70\% Requirements with the Completion Domain Model}
\label{tab:MappingRateInCompletionDomainModel}
\begin{tabular}{l|c|c|c}
\hline
Case & \multicolumn{1}{l|}{\#Entities in REQ} & \multicolumn{1}{l|}{\#Mapped Entities} & \multicolumn{1}{l}{Mapping Rate} \\ \hline
UAV  & 53                                     & 40                                     & 75.47\%                          \\ \hline
BAS  & 145                                    & 104                                    & 71.72\%                            \\ \hline
\end{tabular}
\end{table*}

\textbf{Addressing RQ3:} By aligning the requirements and domain model and completing the domain model with requirements, the mapping rates of domain model and requirements in the two cases are 75.47\% and 71.72\%, increasing by 32.07\% and 51.72\%.

\section{Phase III: Regularity analysis}
\label{subsec: regularityAnalysis}

In this phase, we analyze the distribution regularity of the mapped entities, both from the occurrence in the domain model and requirements. Then we give our initial idea about the missing function recommendation according to the regularity and answer the RQ4.

\subsection{The Distribution of Mapped Entities in Domain Models}
Generally, the mapped entities only account for a small proportion of the domain model because of the specific focus. For example, when implementing the obstacle avoidance function of a UAV, its battery control function may not be considered. Therefore, the recommending scope in the domain model should be narrowed down to more accurately lock the missing functions targeted by the requirements. We analyze the distribution regularity of the mapped entities with the purpose of more accurate recommendations. The regularities are analyzed from the following four aspects.

 \textbf{\emph{The type of entities in domain model}}: As we mentioned in Section \ref{subsubsec:TermsExtraction}, according to the OWL standard, the elements in domain model can be typically divided into multiple categories: \emph{Classes}, \emph{Object Properties}, \emph{Data Properties}, \emph{Named Individuals}, and \emph{Annotation Properties}. \emph{Classes} provide an abstraction mechanism for grouping resources with similar characteristics \cite{bechhofer2004owl}. \emph{Named Individuals} represent the objects in a domain, i.e., instances of Classes. \emph{Properties} (also called as \emph{Object Properties}) represent a kind of relation between individuals. \emph{Data Properties} link individuals to data values. \emph{Annotation Properties}, which is a kind of metadata, can be used to interpret Classes, individuals, object/Data properties. 

 We explore the types of mapped entities in the original domain model and the model after completion and make observations based on the two cases.
    \begin{itemize}
    	\item For UAV:
    	\begin{itemize}
    	    \item In the original domain model, there are 400 entities in total, and 146 ones are the types of \emph{Classes}, accounting for 36.5\%. In the 53 terms from 70 requirements, there are 23 mapped to the domain model entities, 21 of which are Classes (i.e., 91.3\%), and the remaining two are Named Individuals.
    	   \item In the completion domain model, there are 548 entities in total, and 294 ones are with the type of \emph{Classes}, accounting for 53.64\%. In the 137 terms from 70 requirements, 40 can be mapped to the domain model, 38 of which are Classes (i.e., 95\%).
    	   \end{itemize}
    	\item For BAS:
    	\begin{itemize}
    	    \item In the original domain model, there are 484 entities, and all of them are the types of Classes. So, all of the mapped entities are Classes. In the 145 entities in 320 requirements, 29 can be mapped with the entities in the domain model, all of which are Classes.
    	    \item For the completion model, there are 1135 entities, and all of them are types of Classes. All of the 104 mapped entities are Classes.
    	\end{itemize}
    	 
    \end{itemize} 

\begin{center}
\begin{tcolorbox}[colback=gray!10,
                  colframe=black,
                  width=8cm,
                  arc=1mm, auto outer arc,
                  boxrule=0.5pt,
                 ]
        \emph{\textbf{Observation I}: More than 90\% of the mapped entities between software requirements and the open domain models are of the Classes entities in the original domain model. And in the completion domain model, more than 95\% of the mapped entities are of the Classes type.}
\end{tcolorbox}
\end{center}
      
\begin{figure*}[!htbp]
    	\centering
    	\includegraphics[width=1\textwidth]{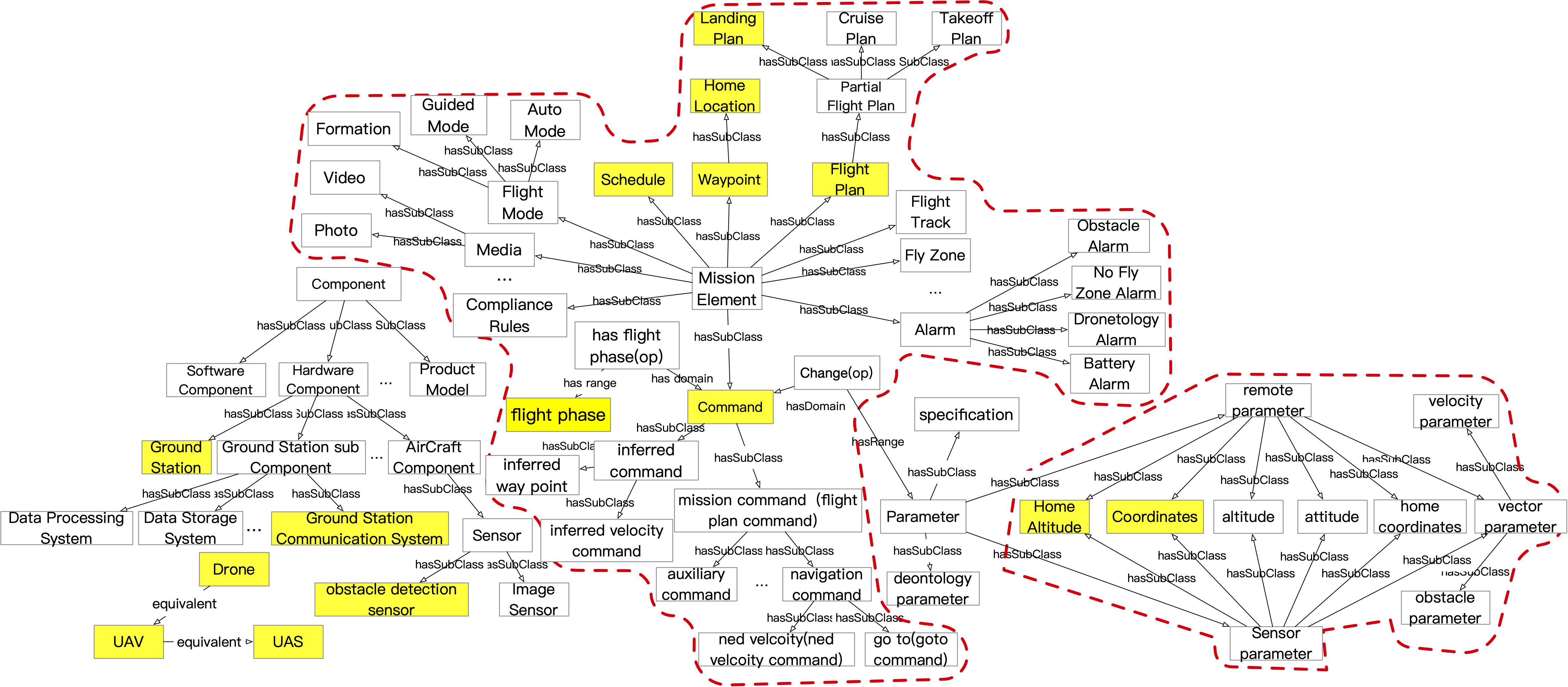}
    	\caption{Distribution of the mapped nodes in the domain model of UAV (The highlighting part with red border \\is the range of the family of UAV domain model calculated by the AHME-based method)}
    	\label{fig:distributionUAV}
    \end{figure*}

\begin{figure*}[!htbp]
    	\centering
    	\centerline{\includegraphics[width=0.9\textwidth]{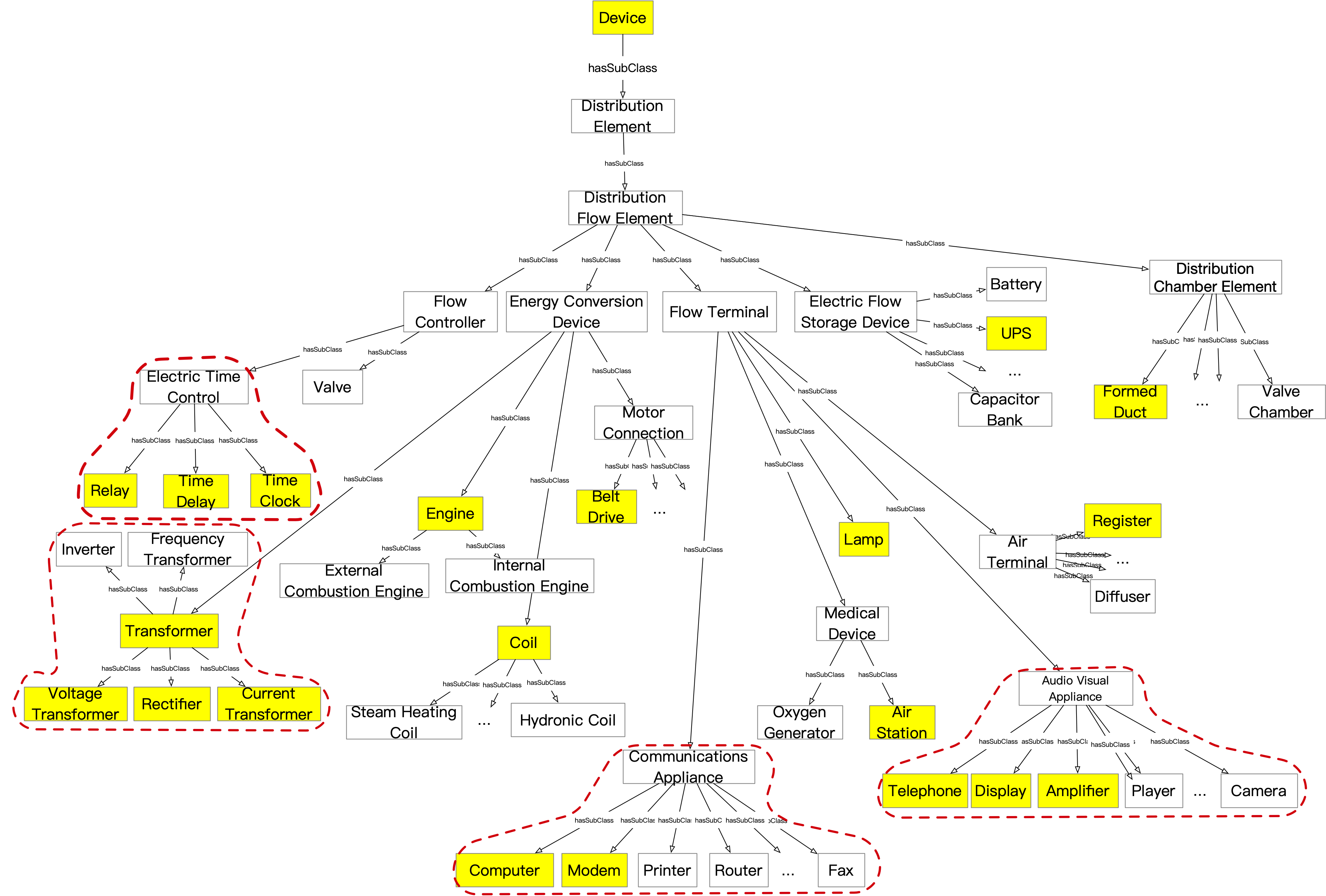}} 
    	\caption{Distribution of the mapped nodes in the domain model of BAS (The highlighting part with red border \\is the range of the family of BAS domain model calculated by the AHME-based method)}
    	\label{fig:distributionBAS}
    \end{figure*}

    \textbf{\emph{The distribution of mapped entities in domain model}}: We attempt to observe the mapping regularity from their distribution on the graph of the domain model. For this purpose, we select and show the sub-trees of the original domain model, which contain the most concentrated mapped entities, highlighted with the yellow background, shown in Fig.~\ref{fig:distributionUAV} and \ref{fig:distributionBAS}.
    
\begin{table*}[htbp]
\centering
  	\caption{The levels of mapped entities in the domain models }
  	\label{tab:nodeType}
    \begin{tabular}{l|c|c|c|c|c}
    \hline
Case                                      & \multicolumn{1}{l|}{Domain Model} & \begin{tabular}[c]{@{}c@{}}Root Node\end{tabular} & \begin{tabular}[c]{@{}c@{}}Intermediate Node\end{tabular} & \begin{tabular}[c]{@{}c@{}}Leaf  Node\end{tabular} & \multicolumn{1}{l}{\begin{tabular}[c]{@{}l@{}}Proportion of Leaf Nodes\end{tabular}} \\ \hline
\multicolumn{1}{c|}{\multirow{2}{*}{UAV}} & Original                          & 2                                                   & 4                                                           & 17                                                  & 73.9\%                                                                                 \\ \cline{2-6} 
\multicolumn{1}{c|}{}                     & Completion                        & 2                                                   & 4                                                           & 48                                                  & 88.89\%                                                                                  \\ \hline
\multirow{2}{*}{BAS}                      & Original                          & 1                                                   & 8                                                           & 21                                                  & 72.4\%                                                                                 \\ \cline{2-6} 
                                          & Completion                        & 1                                                   & 8                                                           & 66                                                  & 88\%                                                                                   \\ \hline
\end{tabular}
\end{table*}

  We can find that the mapped entities tend to be leaf nodes from these two figures. We calculate the ratio of leaf nodes in all mapped entities in the two domain models, shown in Table.~\ref{tab:nodeType}. For the UAV case, in the 23 mapped entities, 17 are leaf nodes, two abstract concept entities, and four intermediate nodes. Leaf nodes account for 73.9\% of all mapped entities. Moreover, 29 mapped entities for the BAS case include 21 leaf nodes, eight intermediate nodes, and one abstract concept entity. Leaf nodes account for 72.4\% of mapped entities.
  
  In the two completion domain models, the ratio of the mapped leaf entities has an obvious improvement (about 15\%). The reason is simple during the model completion, most of the requirement entities were added as the leaf nodes of the domain model (seen in Section \ref{subsec:modelCompletion}).

\begin{center}
\begin{tcolorbox}[colback=gray!10,
                  colframe=black,
                  width=8cm,
                  arc=1mm, auto outer arc,
                  boxrule=0.5pt,
                 ]
    \emph{\textbf{Observation II}: More than 70\% mapped entities are leaf nodes in the trees of domain models. Besides, the terms of software requirements usually appear in a few sub-trees, reflecting specific focuses. And more than 85\% mapped entities are leaf nodes in the trees of our two completion domain models.}
\end{tcolorbox}
\end{center}

 \textbf{\emph{The family belonging of the mapped entities in domain model}}: From Fig.~\ref{fig:distributionUAV} and \ref{fig:distributionBAS}, we can also see that the distribution scope of the mapped entities in the domain model tends to be concentrated in multiple child nodes under one or several intermediate nodes. 
 This phenomenon is in accordance with our hypothesis that there is always a few particular focuses on the requirements of a specific version of software products rather than the entire domain. This observation inspires us first to capture the focus of the existing software requirements when recommending the missing functional points. With these belongings, we can significantly reduce the search scope of missing information. For example, in Fig.~\ref{fig:distributionUAV}, there are 46 entities in the red dashed box, while the number of class entities in the entire UAV domain model is 146, accounting for 31.5\% of the total. In Fig.~\ref{fig:distributionBAS}, the number of entities contained in the red dashed box is 33, and the number of class entities in the BAS domain model is 484, accounting for 6.82\%. In other words, the belonging scopes are helpful for the specific missing information retrieval.

Domain models typically associate parent and child nodes in the form of the tree structure. However, while performing further analysis, we can see no direct parent-children relationship between many mapped entities. Still, they probably have the same parent or ancestor (e.g., \emph{Actuator} and \emph{Alarm}). To explore the focuses of software requirements, finding a common ancestor belonging to the most mapped entities is necessary. This ancestor should be far from the root, and its level should be as high as possible (to be more specific). We want to recommend the missing entities in the tree with the root of this ancestor. Thus, we propose a metric AHME (the Ancestors of the Highest level with most Mapped entities) for locating the entity that can be the parent or ancestor of the most mapped entities and is at the highest level of the domain model. It can be defined formula as follows.

\par
\begin{equation}
AHME= \frac{MED}{ME} * LEVEL(node)
\end{equation}

\emph{MED} indicates the number of mapped entities belonging to the descendants of one node, \emph{ME} is the total amount of mapped entities, and \emph{LEVEL(node)} indicates the level of this node. 
We calculate the AHME value for each node in the domain model, select one or more of the highest AHME values (not all mapped entities will be included), and take that node and all its children as the scope of the requirements in the domain model. 

We use the metric of AHME to locate the family belonging to the requirements of UAV and BAS in their corresponding domain model.

For the original UAV domain model, we first calculated the AHME values for all \emph{Classes} entities and found that the AHME values of \emph{MissionElement}, \emph{RemoteParameter}, and \emph{SensorParameter} were much higher than other Entities. The AHME value of the \emph{MissionElement} is 0.26. Its height is 1, and there are six mapped entities in its children nodes. And the \emph{RemoteParameter} and \emph{SensorParameter} have an AHME value of 0.43, both with a height of 5, and two mapped entities in their children nodes. We concluded that the information of this requirement is mainly in the range of these three families, with the red circle in the mapped nodes distribution graph in Fig.~\ref{fig:distributionUAV}. 

Then we use the same method to analyze the AHME values of the mapped nodes in the BAS case and find that the AHME values of \emph{ElectricTimeControl}, \emph{Transformer}, \emph{CommunicationsAppliance} and \emph{AudioVisualAppliance} (highlighted with the red border in Fig.~\ref{fig:distributionBAS}) is higher than other entities, reflecting the overlapped concerns. 

 \begin{center}
\begin{tcolorbox}[colback=gray!10,
                  colframe=black,
                  width=8cm,
                  arc=1mm, auto outer arc,
                  boxrule=0.5pt,
                 ]
 \emph{\textbf{Observation III}: The mapped entities tend to be contained in a few families of nodes in the open domain model, and the root node of these families reflects their focuses to a certain extent.}
 
\end{tcolorbox}
\end{center}

 \subsection{The distribution of Mapped Entities in Requirements}
 
 The mapped entities in software requirement descriptions usually have some semantic relationships (for example, the subject and object are connected by the predicate). However, there may be no direct connection between them in the domain model but the indirect path with a few hops. This difference makes the entity-relationship-based recommendation difficult because the search complexity in the worst case is exponential ($n\times(n-1)C_p^2$, which assumes that there are \emph{n} entities in the domain model, and the entities in \emph{p} hops are considered), let alone the computation complexity. Text embedding technologies may help address this problem. In this study, we observe the distribution of mapped entities in requirements and expect to find factual support for this sort of technology. 
 
 Our analysis is performed from three aspects: 1) the proportion of requirements including mapped entities in total requirements; 2) the proportion of requirements containing two or more mapped entities; 3) the kinds of dependencies between two or more mapped entities in single requirements.
 
\emph{\textbf{The proportion of requirements including mapped entities.}} In 70\% of the UAV requirements (i.e.,70), 51 requirements contained mapped entities, accounting for 72.86\%. In 70\% of the BAS set (i.e.,320), 180 requirements contain mapped entities, accounting for 56.25\%. These values reflect the matching degree of the software requirements and the domain model. The higher the values, the better of the matching degree, and the more effective recommendation based on the model.

\emph{\textbf{The proportion of requirements containing two or more mapped entities.}} Of the 51 requirements containing mapped entities of the 70\% requirements of UAV, 35 contain at least two mapped entities, accounting for 68.63\%. For the requirements of BAS, this ratio is 31.25\%. In order to find the reason for the big difference between the result of the 2 cases, we analyzed the data sets of UAV and BAS and found that the interaction between entities in requirements was higher in UAV than in BAS. In BAS, the states of many entities are described rather than the interaction between entities.

\emph{\textbf{Relationship types of mapped entities in requirements.}} To obtain the relationship types of the mapped entities in requirements, we walked through the requirements containing two or more entities in the two sets. We found that most of the relationship is subject-object (about 97.14\% in the 35 requirements of UAV, and about 99\% in the 100 requirements of BAS), and some juxtaposition relationship (AND or OR) exist too (about 11.43\% in the 35 requirements of UAV, and about 44\% in the 146 requirements of BAS). 
These statistics of relationship types are expected to improve the missing concept recommendation in future.

  \begin{center}
\begin{tcolorbox}[colback=gray!10,
                  colframe=black,
                  width=8cm,
                  arc=1mm, auto outer arc,
                  boxrule=0.5pt,
                 ]
\emph{\textbf{Observation IV}: More than 60\% of the requirements contain mapping entities, of which 50\% of requirements contain multiple mapped entities. The entities are usually in (subject-object) relationships, and part is juxtaposition relationships connected by (AND, OR).}
 
\end{tcolorbox}
\end{center}
 
\textbf{Addressing RQ4:} By analyzing the mapped terms between requirements and the open domain models in two case domains, we made four observations (i.e., regularities) from the entity type, node distribution, family belonging in the domain model, and three frequency-related characteristics in requirement statements. The analysis of two case domains illustrates that these regularities may potentially help recommend and reduce the searching scope of the missing information in requirements regarding the plumbline of the open domain model. 

\section{Verification: Usefulness of the Regularities}
\label{sec:usefulness}

To answer RQ5 and RQ6, we design an experiment. We regard the remaining 30\% requirements of the two sets as the missing ones and recommend them according to the 70\% requirements and the open domain model. We evaluate the usefulness of these regularities by calculating the extent to which the concepts in the 30\% requirements can be recommended correctly, with the common metrics of Recall, Precision, and $F_{2}$ because the recall is more important \cite{berry2017evaluation}. Recall measures the extent to which the correct missing information can be identified automatically. Precision measures the ratio of correct information about missing requirements in all automated recommendations. Considering the stochastic overlap between the selected and remaining requirements, we run the experiments 30 times to obtain the average values of the metrics.

In this study, we only evaluate the effectiveness of the distribution regularities in the (original and completion) domain model, and those in requirements will be explored in the future. 

The measurement results on the original domain model are shown in Table~ \ref{tab:regularityResult}. In this table, we evaluate the effectiveness of each single regularity and the combination of them in the two domains. For the convenience of comparisons, we also calculate the three metrics of recommendations without using any of our regularities, shown in the first row. From the columns of $F_2$, we can see that our regularities are indeed helpful for the missing information recommendation (with better $F_2$ than that without regularities). In addition, we can also see that the recall values tend to decrease from top to bottom. In other words, most of the concepts in the missing 30\% requirements can be mapped to the \emph{Classes} entities (87.5\% for UAV and 100\% for BAS). However, most of the \emph{Classes} entities cannot be mapped to the concepts in requirements (i.e., the \emph{Precision} values are very low) due to the pretty sparse mapping. Thus we need more strong strategies to help reduce the scope of recommendations. Similar results can be obtained from the row of \emph{node type}.

Luckily, we can see that the \emph{family belonging} based on our AHME is the most effective regularity yielding the best recall and precision values. Considering the overlapping ratio between the 70\% requirements and the original domain model (i.e., 43.4\% for UAV and 20\% for BAS seen in Table.~\ref{tab:MappingRateInDifferentApp}), we think the recommendation based on the AHME is reasonable. It also reflects that as parts of the requirements of one specific software system, both the missing and existing requirements focus on the same group of concerns. One possible reason is that we split the whole existing requirements into two parts, and for the ``\emph{real}'' missing requirements recommended in practice, the phenomenon may be different, which needs further exploration.

\begin{table*}[htbp]
    \centering
    \caption{The effectiveness of our regularities  for missing information recommendation in the two cases (Original Domain Model)}
    \label{tab:regularityResult}
    \begin{tabular}{|c|c c c|ccc|}
        \hline
       \multirow{2}{*}{Regularity}  &\multicolumn{3}{c|}{UAV} &\multicolumn{3}{c|}{BAS} \\ \cline{2-7}
       & Recall & Precision & $F_2$ & Recall & Precision & $F_2$ \\ \hline
        Without Regularities   & 1.0 & 0.03 & 0.13  & 1.0 & 0.04 & 0.17  \\ \hline 
        Entity Type  & 0.875 & 0.112 & 0.37  & 1.0  & 0.071 & 0.28   \\ \hline
        Node Type & 0.7 & 0.07 & 0.25 & 1.0  & 0.091 & 0.33  \\ \hline
        Family Belonging  & 0.32 & 0.16 & 0.26 & 0.5   & 0.24 & 0.41 \\ \hline
        \begin{tabular}[c]{@{}c@{}}Regularity Combination\end{tabular} & 0.21 & 0.18  & 0.20  & 0.5   & 0.24 & 0.41  \\ \hline
    \end{tabular}
\end{table*}

Due to the most effective of the family belonging regularity seen from the Table.~\ref{tab:regularityResult}, we further measure to what extent we can recommend the missing information \emph{that should be in the family zone} according to this regularity with the recall, Precision and $F_2$, shown in Table.~\ref{tab:FamilyCaseResult}. Recall measures the ratio of concepts we correctly identified as the family belonging, and Precision measures the extent of correct recommendations according to the family belonging. We can see that the recommendations are much better than the results in Table.~\ref{tab:regularityResult}. Particularly, almost all of the entities, which should be recommended, can be identified, although the mapping is sparse and the entity number is therefore small. 

From Table.~\ref{tab:regularityResult}, we can also see that the simple combination of the regularities does not make a significant improvement on the recommendation, although the Precision gets slightly better than that of the pure \emph{Family Belonging} (seen in Table.~\ref{tab:regularityResult}). This triggers a better combination approach in the future.

\begin{table*}[htbp]
    \centering
    \caption{The recommendation effectiveness with the scope of Family Belonging in the original domain model}
    \label{tab:FamilyCaseResult}
    \begin{tabular}{c|c c c c}
        \hline
        \multicolumn{5}{c}{UAV} \\ \hline
        Family Belonging & \begin{tabular}[c]{@{}c@{}}Actually/Should have\end{tabular} & Recall & Precision & $F_2$ \\ \hline  
        Mission Element  & 6/7                & 0.857  & 0.18      & 0.49    \\ 
        Remote Parameter & 2/2                   & 1.0      & 0.25      & 0.63   \\ 
        Sensor Parameter & 2/2   & 1      & 0.25      & 0.63     \\ \hline
        Average    &    & 0.9    & 0.21      & 0.54  \\ \hline
        \multicolumn{5}{c}{BAS} \\ \hline
        Family Belonging         & \begin{tabular}[c]{@{}c@{}}Actually/Should have\end{tabular} & Recall & Precision & $F_2$ \\ \hline
        Electric Time Control   & 3/3                  & 1.0      & 0.75      & 0.94    \\
        Transformer    & 3/3                 & 1.0      & 0.67      & 0.91                                          \\ 
        Communications Appliance & 2/3                & 0.67   & 0.17      & 0.42   \\ 
        Audio Visual Appliance   & 3/3                 & 1.0      & 0.27      & 0.65   \\  \hline
        Average    &      & 0.91   & 0.37      & 0.70                                          \\ \hline
\end{tabular}
\end{table*}

In order to evaluate the effectiveness of the completion model on missing requirement entities recommendation (addressing RQ6), due to the best performance of the pure family belonging regularity, we perform the recommendation on it, shown in Table.~\ref{tab:recommendationComparison}. With the convenience of comparisons, we also compute the gains of it on the three metrics than the recommendation with the AHME family belonging to the original domain model.

Due to the expansion of the domain model, more entities are involved. Due to the unknown associations between the 70\% requirements and the remaining 30\% ones, this probably brings more noise for the recommendation. In addition, it is possible that the open domain model can't cover the remaining 30\% requirements, even after completion. This means that more false-positive results would be most likely to be recommended. Thus, one potential problem of the domain model completion for the missing recommendation is the lower precision. However, from Table.~\ref{tab:recommendationComparison}, we can see that the improved approach obviously increases the recall (with the gains of 41\% for UAV and 72\% for BAS)  with almost no loss of precision (2\% for UAV and 1\% for BAS). After analysis, we find that the 70\% requirements indeed extend the family belongings, and meanwhile, the added entities provide more strong hints about the entities in the missing requirements.

Besides comparing the recommendation with our full approach (i.e., family belonging in the extended domain model) in Table.~\ref{tab:recommendationComparison} and the recommendation without any regularity in the original domain model in Table.~\ref{tab:regularityResult}, we can find that our approach yields the $F_2$ gains of 146\% in UAV and 223\% in BAS. This is strong proof of the effectiveness of our approach in missing requirements detection based on the open domain model.

In order to further illustrate the usefulness of domain model completion on the missing requirements recommendation, we analyze the source of the right recommended entities. We find that for the UAV case, about 23.06\% right recommendations are from the original domain model, and the remaining 76.94\% are from the extended part. In the BAS case, about 44.12\% right recommended entities are from the original domain model, and the other 55.88\% are from the extended part. These detailed numbers also provide proof about the effectiveness of requirement associations (although uncertain) to missing requirements recommendations.

\begin{table*}[htbp]
    \centering
    \caption{The comparison of the recommendation effectiveness based on family belonging}
    \label{tab:recommendationComparison}
    \begin{tabular}{|c|c c c|ccc|}
        \hline
       \multirow{2}{*}{Domain model}  &\multicolumn{3}{c|}{UAV} &\multicolumn{3}{c|}{BAS} \\ \cline{2-7}
       & R. & P. & $F_2$ & R. & P. & $F_2$ \\ \hline
        original model & 0.32 & 0.16 & 0.26 & 0.5  & 0.24 & 0.41  \\ \hline
        completion model & 0.45 & 0.14 & 0.32 & 0.86   & 0.23 & 0.55 \\ \hline
        Gain & 0.41 & -0.12  & 0.23   & 0.72 & -0.04 &0.34 \\ \hline
    \end{tabular}  
\end{table*}

\textbf{Addressing RQ5 and RQ6:} Our regularities are indeed helpful for the missing requirements recommendation, with the $F_2$ increasing of 13\%-24\% than that without any regularities. And the completion domain model, especially with the AHME regularity, can obviously help improve the recommendation with the $F_2$ gains of 23\% and 34\% in the two cases, than that with the original domain model using the same regularity. What's more, using the AHME regularity in the completion domain model can reach obvious outperformance than the recommendation with no regularity in the original domain model, with the $F_2$ gains of 146\% and 233\%.

\section{Discussion}
\label{sec:discussion}

In this section, we discuss the threats to validity, the implications and the limitations of this study.

\subsection{Threats to Validity}

\subsubsection{\textbf{Internal Validity}} The main threat to the internal validity of our work is from the perspectives of regularities we select. As a preliminary study, we only analyze the regularities from four relatively intuitive perspectives to explore the usefulness of the open domain model in the missing requirements recommendation. Undoubtedly, there must be other analytical perspectives, such as the relevance between mapped and unmapped entities and their semantic dependencies. We would like to dig them in future.

Another threat to internal validity is from the automated domain model completion. We improved the existing approach of TransE \cite{zhong2015aligning} for the model alignment and completed the model based on angles between entity vectors. This process is stochastic to a certain degree. To mitigate the impact on the final results, we run the whole process, including the model completion and missing recommendation 30 times, to obtain the average values of the metrics.
 
 \subsubsection{\textbf{External Validity}} This study is performed based on two distinct case domains. Both the domain model and requirement descriptions are from different groups, thus providing confidence in the external validity of our findings to a certain degree. That being said, we emphasize that our findings are based on matching domain models and requirements. As long as there is a certain degree of overlap, the domain model is helpful to the improvement of the requirement completeness. Moreover, the usefulness degree depends on the overlap ratio and also the strategies of recommendation. 
 
\subsubsection{\textbf{Conclusion Validity}} We evaluate the effectiveness of the regularities by comparing our recommendations with the concepts in the 30\% simulating missing requirements. The results show that the regularities can indeed reduce the searching scope and increase the accuracy. However, due to the very specific focus of requirements and the continuous evolution of systems, the recommendations uncovered by the 30\% requirements may be valid too. This means our results may have some potential effects which need further validation.
 
  \subsection{Implications}
 There are three lessons we learned from this study.
 
 1) It is essential to build the mapping between the open domain model and requirements in this work, and the mapping results will seriously affect the missing information identification. We used a hybrid of term-based direct and synonymous-based indirect mapping, and about 79\%-90.0\% mapping can be established correctly. Although the synonymous detection is complex and requires more domain corpus, it is worthwhile.
 
 2) Undoubtedly, the domain model completion spends more time (i.e., for UAV, 20 minutes once; for BAS, 90 minutes once). Looking at the improvements on the recommendation (i.e., 23\% and 34\% $F_2$ gains), we think this process makes sense. Besides, we only show the effectiveness of the completion model on the missing requirements recommendation, not the correctness or domain compliance. This may need more exploration in the future.
 
 3) We attempt to reduce the searching scope of the missing information in domain models according to the four regularities. However, we did not give specific recommending algorithms. \emph{How to recommend the domain model entities as missing information is a problem worthy of research}.

\subsection{Limitations}
In general, there are four primary limitations in this study.

\emph{\textbf{The elements of domain model under analysis is limited.}} We only build the mapping between the \emph{Classes} entities in the domain model and the concepts in requirement descriptions and didn't consider other types of elements of the domain model, such as attributes of concepts. However, Arora et al. showed that all of the entities, attributes, and their associations of the requirement-based domain model are sensitive to the missing requirements \cite{LionelESEM}. We would like to explore the usefulness of other parts in the future.

\emph{\textbf{Unmapped entities are not involved in our analysis.}} In the two cases, unmapped entities account for a large proportion of both domain models and software requirements. To recommend the missing requirements, we focus on the mapped entities because they illustrate the picture of \emph{what we have already}. However, the unmapped entities work as the necessary context for these mapped entities. Therefore, it is reasonable to expect better recommendations by considering the semantical relationship between the mapped and unmapped entities.

\emph{\textbf{The strategies for recommendation can be improved.}} This preliminary study focuses on exploring the answer of \emph{yes} or \emph{no} to whether the open domain model is useful for missing requirements detection. Thus, the regularities we propose aim to reduce the scope of recommendation with sparse mapping considerations. Besides the distributions of the mapped entities in the domain model or requirements, lots of other semantical information, such as the relevance between the entities \cite{zhu2016computing,manda2018statistical}, can be used. We will explore this in the future.

\section{Conclusion}
\label{sec:conclusion}

The missing (critical) software requirements may lead to disastrous consequences.
Lots of research validate the completeness of software requirements based on requirements oriented domain models, which usually do not exist for most domains. Fortunately, it is not difficult to find domain models online constructed for variable purposes for various domains. Thus we explore the usefulness of these open domain models on the missing requirements recommendation via two case domains, whose requirements and domain models are from different groups. We proposed to establish the mapping between requirements and the open domain model and to complete the domain model with the known requirements with the purpose of closing down the gap between them. Also, we observed four regularities of the mapped entities to help reduce the searching scope of missing information in requirements to prove their effectiveness. Although their usefulness is relatively limited, surprisingly, it helps discover missing information in the requirements, especially the missing concerns. Thus, we can say that the open domain model can be potentially used as an effective plumbline for recommending the missing functional information. We would like to explore the automated recommendation approach in the future.

\section{Acknowledgment} 
*** *** *** *** *** *** *** *** *** *** *** ***


\bibliographystyle{plain}
\bibliography{Reference}

\end{document}